\documentclass{emulateapj}

\def\etal{{\it et al.}}
\def\kms{km~s$^{-1}$}

\def\msun{$M_\odot$}

\begin{document}
%\hskip 3.5in{Version 4.0\hskip 10pt 07Jun2007}
\title{NGC~4254: An Act of Harassment Uncovered by the Arecibo Legacy Fast ALFA Survey}

\author{Martha P. Haynes\altaffilmark{1,2},Riccardo Giovanelli\altaffilmark{1,2}, 
 and Brian R. Kent\altaffilmark{1}}

\altaffiltext{1}{Center for Radiophysics and Space Research, Space Sciences Building,
Cornell University, Ithaca, NY 14853. {\it e--mail:}haynes@astro.cornell.edu, 
riccardo@astro.cornell.edu, bkent@astro.cornell.edu}

\altaffiltext{2}{National Astronomy and Ionosphere Center, Cornell University,
Space Sciences Building,
Ithaca, NY 14853. The National Astronomy and Ionosphere Center is operated
by Cornell University under a cooperative agreement with the National Science
Foundation.}

\begin{abstract}
We present an HI map constructed from the Arecibo Legacy
Fast ALFA survey of the surroundings of the strongly asymmetric
Virgo cluster Sc galaxy NGC~4254. Noted previously for its lopsided appearance,
rich interstellar medium, and extradisk HI emission, NGC~4254 is believed 
to be entering the Virgo environment for the first time and at high
speed. The ALFALFA map clearly shows a long HI tail extending $\sim 250$ kpc 
northward from the galaxy. Embedded as one condensation within this HI structure 
is the object previously identified as a ``dark galaxy'':  Virgo~HI21 (Davies \etal~2004).
A body of evidence including its location within and velocity with
respect to the cluster and the appearance and kinematics of its strong 
spiral pattern, extra-disk HI and lengthy HI tail is consistent with a
picture of ``galaxy harassment'' as proposed by Moore \etal ~(1996a,b; 1998).
The smoothly varying radial velocity field along the tail as it emerges
from NGC~4254 can be used as a timing tool, if interpreted as resulting 
from the coupling of the rotation of the disk and the collective gravitational 
forces associated with the harassment mechanism.
\end{abstract}


\section {Introduction}\label{intro}

Because of its proximity and relative richness, the Virgo cluster 
provides an especially useful laboratory for witnessing in detail 
the impact of environment on galaxy evolution.  In Virgo, 
studies of the HI deficiency first quantified the impact of the cluster 
environment (Davies \& Lewis 1973; Chamaraux, Balkowski \& Gerard 1980), 
while subsequent mapping of the gas distribution proved that the HI 
disks of the highly gas-poor objects were systematically smaller than 
those of galaxies of normal HI content (Giovanelli \& Haynes 1983; 
Cayatte \etal ~1990). Further HI synthesis observations, 
most recently those of the VLA Imaging of Virgo Galaxies (VIVA)
Survey (Chung \etal ~2007), yield details of the HI distribution in a 
growing selection of Virgo galaxies. Concurrently, the Arecibo Legacy Fast 
ALFA (ALFALFA) survey is conducting a sensitive blind HI survey of the 
entire Virgo cluster region, identifying a host of HI features that are 
not coincident with stellar counterparts (Kent \etal ~2007). The combined 
picture provided by both the blind and targeted HI surveys of Virgo promise 
to allow evaluation of the relative importance of interaction processes 
likely at work in the Virgo environment due to both gravitational and
hydrodynamic forces.

NGC~4254 (= M99) is a bright Sc galaxy located 3.7$^\circ$ to the NW of M87.
Binggeli \etal ~(1985) assigned it membership in the main Virgo A cluster, so
that its projected separation from M87 is $\sim$1 Mpc for a Virgo distance 
of 16.7 Mpc (Mei \etal ~2007).
Its heliocentric velocity of 2404 \kms ~implies a velocity of $\sim$ 1300 \kms
~with respect to the cluster overall. 
It has no close neighbors but is noted for its strong
$m=1$ asymmetric spiral pattern and vigorous star formation, especially
in its bright southern arm. NGC~4254 has been the subject of numerous 
detailed studies, most of which focus on the generation of its unusual 
optical lopsidedness by interaction with the cluster environment.

Of relevance to the present discussion,
Davies \etal ~(2004) reported the detection, later
confirmed via corroborating Arecibo observations (Minchin \etal ~2005a),
of 2.2 $\times$ 10$^8$ \msun ~of HI at a position some 120 kpc north
of NGC~4254. HI emission was detected in a total of five adjacent 
Arecibo pointings, spaced at intervals of 2.7\arcmin, and 
centered on ($\alpha, \delta$, J2000) 12$^h$17$^m$53.6$^s$,
+14$^\circ$45\arcmin25\arcsec. 
Based on the Arecibo observations, especially the relatively broad HI line 
width of 220 \kms, and the lack of detectable HI in a VLA synthesis
map, they argued that Virgo~HI21 is the disk of a large ($>$16 kpc 
in diameter), optically--dark galaxy with a much larger total mass than
that of the HI itself (Minchin \etal~ 2005b). They noted that Virgo~HI21 
lies about 120 kpc from both NGC~4262 and NGC~4254 but argued against 
any tidal connection with either, both because of the relatively large 
separation and their high relative velocity. 

In this Letter, we exploit the wide field HI maps of 
the ALFALFA survey to detail a long complex of HI clouds, extending
from NGC~4254 northward. Virgo~HI21 is shown to be one condensation
within this HI tail.  Deeper observations subsequently undertaken at 
Arecibo allow us to trace the HI to a distance of $\sim 250$ kpc to 
the north of NGC~4254. In section 2 we describe the ALFALFA 
data and followup Arecibo observations. In section 3, we discuss
the body of evidence regarding NGC~4254 in terms of the galaxy harassment 
scenario of Moore \etal ~(1996a,b; 1998). All coordinates are 
for epoch J2000.0.

\section{Observations and Analysis}\label{obs}

The region around NGC~4254 and Virgo~HI21 has been mapped as part 
of the ALFALFA survey program. ALFALFA is a two-pass survey which uses 
the the Arecibo L-band Feed Array (ALFA) on the 305~m antenna. The 
ALFALFA observing strategy and data processing pipeline were outlined 
in Giovanelli \etal ~(2005). Once both passes of the survey are 
completed, 3-D spectral line data cubes are constructed and a signal 
extraction algorithm, discussed in Saintonge (2007), is applied for
preliminary identification of HI sources. HI sources in the northern 
part of Virgo were included in the first 
ALFALFA catalog release (Giovanelli \etal ~2007), covering between 
$+12^\circ < \delta < +16^\circ$. That tabulation includes 17 
HI detections within 1$^\circ$ of Virgo~HI21 
($\alpha$ = 12$^h$18$^m$, 
$\delta$ = +14$^\circ$44\arcmin) and at velocities $cz < 3000$ \kms.
Of the 17 HI sources, four have no optical counterparts; three of those 
are extended (in comparison with the $3.3'\times 3.8'$ Arecibo telescope
beam) and appear associated with the NGC~4254--Virgo~HI21 system, 
while the remaining one is probably a foreground Milky Way high velocity cloud. 

The left panel of Figure 1 shows the integrated HI flux density 
contours as detected in the ALFALFA data set overlayed on the 2nd generation 
Digital Sky Survey Blue image in the vicinity north of NGC 4254.
The HI emission is integrated over the range $cz$ = 1946 to 2259 \kms ~for the stream and 
$cz$ = 2259 to 2621 for the galaxy on a grid with 1\arcmin ~spacing.
The velocity cut for the stream is outside the 2250-2510 \kms ~range reported 
in VLA maps of NGC 4254 (Vollmer \etal ~2005). The superposed circle at  
$\alpha=$ 12$^h$17$^m$53.6$^s$, $\delta=$ +14$^{\circ}$45$^{\prime}$25$^{\prime\prime}$ 
indicates the centroid of Virgo~HI21 as reported by Minchin \etal ~(2005a).
HI features connect in a stream that appears to emanate from NGC~4254.

In order to confirm the extended nature of the stream and, with higher
signal--to--noise, explore the velocity field along it,
a set of HI spectra were obtained along the ridge of the HI stream with 
longer dwell times than those of the ALFALFA survey.
Thirty-six on/off pairs were taken at 1.5\arcmin ~steps in declination,
centered on the peak emission indicated by the ALFALFA data, using 
the single-beam L-band wide (LBW) receiver of the Arecibo telescope.
Difference spectra were then constructed from each pair, with 2048 channels 
per polarization covering a 25 MHz band centered at 1410 MHz with spectral
resolution of 12.2 kHz (2.6 \kms ~at $cz$ = 2000 \kms). The  
integration times varied from three to 
six minutes on source, depending on the expected signal strength, yielding an 
rms noise per channel of better than $\sim$2.6 mJy before any spectral
smoothing. These observations were about twice as sensitive as those 
obtained through ALFALFA.

The locations of the LBW pointings are illustrated by the filled dots
superposed along the peak of the stream illustrated in the left
panel of Figure 1. The peak velocity at each pointing 
for which the detected signal is significant is indicated on the position-velocity
diagram in the right hand inset. Emission was detected in 27 of the 36 
positions targeted by the LBW observations.
The stream clearly connects to NGC~4254 in velocity space. 

Giovanelli \etal ~(2007) give an HI mass for 
NGC~4254 itself of $4.3 \times 10^9$ \msun. The total HI line flux 
associated with gas in the extended stream, including the clumps reported
by Giovanelli \etal ~(2007) and lesser ones near the ALFALFA detection 
limit is 6.5 Jy-\kms ~which corresponds to M$_{HI} \sim 4.3 \times 10^8$ 
\msun. Given the extended nature of the source, the ALFALFA observations 
miss detecting any diffuse emission at column densities below the 
ALFALFA limit. Assuming that diffuse emission is present at column densities
equal to or less than $3\times 10^{18}$ HI atoms cm$^{-2}$ over a solid angle
of 200 (arcmin)$^2$, up to an additional $1.1\times 10^8$ \msun~ of HI may
be present in the stream, bringing the total HI mass to $5.0\pm0.6\times 10^8$
\msun.
The HI tail can be traced over $\sim$50\arcmin, which at the Virgo distance
corresponds to 242 kpc. This length is
certainly dramatic, but not unique. Of comparable length is the longest
of the stellar streams in the extended halo of M87 
which was detected in very deep images of Virgo by Mihos \etal ~(2005).

\section{Discussion}\label{disc}

By many measures, NGC~4254 is an exceptional Sc galaxy, 
one of the brightest spirals in the Virgo A cluster. Its rare 
dominant m=1 spiral mode is evident in its conspicuous southwestern spiral arm 
and the lesser but still prominent arm with several muted
branches seen to the north and east. There is no primary distance 
measurement but Solanes \etal ~(2002) give a mean 
distance modulus  of 31.04 $\pm$ 0.04, derived from several applications 
of the Tully--Fisher relation, placing it very close to the
adopted cluster distance but at a projected separation of 1 Mpc from
M87. For 35 galaxies with known redshift within 
a projected separation of NGC~4254 of 300 kpc, the mean heliocentric 
velocity is 823 \kms, with a  dispersion of 758 \kms. With a heliocentric
velocity of 2404 \kms ~(Giovanelli \etal ~2007), NGC~4254 is 
moving at high speed with respect to the cluster and most of its
projected neighbors, and is, in fact, the
object with the highest radial velocity seen in that portion of the cluster.

Because of its brightness, size, high gas content and optical asymmetry, 
NGC~4254 has been the subject of a number of detailed investigations whose principal
aim has been to understand the origin of its lopsidedness and strong $m=1$ spiral mode,
particularly in the absence of an appreciable bar or nearby companion. Various
studies of the gaseous components have confirmed the strongly
asymmetric optical structure. The H$\alpha$ (Phookun, Vogel \& Mundy 1993; 
Chemin \etal ~2006) and $^{12}$CO(J=1-0) maps are consistent with the picture of an
externally--forced spiral density wave. Explanations
for its asymmetry include the superposition of spiral modes induced
by global gravitational instability (Iye \etal ~1982), 
the asymmetric accretion of gas onto the disk (Bournard \etal ~2005), ram pressure
stripping (Phookun, Vogel \& Mundy 1993; Sofue \etal ~2003) and a close
high speed encounter plus ram pressure (Vollmer, Huchtmeier \& van Driel 2005).

In contrast to many spirals found within the inner 1.5 Mpc of Virgo, NGC~4254 appears 
to have a normal HI content relative to field Sc's of comparable linear size, a
healthy H$\alpha$ extent relative to its older R-band stellar population 
(Koopmann, Haynes \& Catinella 2006), a relatively normal molecular content 
(Nakanishi \etal ~2006), and a typical disk metallicity gradient(Zaritsky 
\etal ~1994). Hence, NGC~4254 appears not to have yet suffered significant
gas removal or star formation quenching.

Using VLA HI synthesis mapping, Phookun, Vogel \& Mundy (1993) first 
reported the presence of an extra component
of HI superposed on the main, rotating disk of NGC~4254.
They estimated the HI mass of the disk to be $\sim$ 
4.8 $\times$ 10$^9$ \msun~ with the mass in 
``high velocity clouds'' $\sim$ 3\% of that number. As seen in their VLA
HI map (Figure 5 of Phookun, Vogel \& Mundy 1993), the extra gas forms
a loop on the NE side of the galaxy at high velocity, complemented by several
lower velocity clouds which form an extended tail on the SW side. The
latter connect to the much longer HI stream seen in the
ALFALFA data. This connection is also clearly evident in the top panel 
of Figure 1 of Vollmer, Huchtmeier \& van Driel (2005) who reprocessed the
original VLA map. 

Details of the distortion of the HI layer in a number of galaxies
in the Virgo cluster are seen in an accumulating number of HI
maps of Virgo galaxies. The most dramatic evidence of gas compression
and stripping is seen in the HI deficient spirals found close to the
cluster center, yet extra-disk gas is found in a variety of cases.
Oosterloo \& van Gorkom (2005) propose that an HI cloud of 
3.4 $\times$ 10$^8$ \msun~ and extending some 100 by 25 kpc has
been stripped from the strongly deficient Virgo core spiral NGC~4388
by ram pressure due to the surrounding intracluster medium.
Further out from M87, Chung \etal ~(2007) report several 
one--sided HI tails pointing away from M87 and suggest they result from
ram pressure exerted by the hot intracluster medium.

NGC~4254 is moving at high velocity with respect to the cluster but
is located 1 Mpc from M87 where the intracluster medium density is quite low.
While Sofue \etal ~(2003) conclude
that the distribution and kinematics of the molecular gas in the
inner regions of the galaxy are consistent with the effects of
ram pressure as NGC~4254 falls into the Virgo cluster,
Cayatte \etal ~(1994) suggest that the gravitational restoring force
within NGC~4254 probably exceeds the ram pressure force at its
current location on the outskirts of the X-ray emitting region. That ram pressure
alone cannot explain the HI and optical asymmetry is further emphasized by 
Vollmer, Huchtmeier \& van Driel (2005).

While ram pressure is often invoked to explain lopsidedness, gas compression and 
HI deficiency, gravitational rather than hydrodynamical forces may produce
similar effects. Of particular relevance are the simulations of galaxy
harassment reported by Moore \etal ~(1996a) and Moore, 
Lake \& Katz (1998). Those authors explored the collective impact of both high 
speed encounters and the tidal effects of a smooth cluster potential in
a variety of cases, one of which was a spiral galaxy with circular
velocity $\sim$160 \kms, nearly identical to that of NGC~4254. 
The most outstanding characteristics of the HI tail reported here are 
consistent with the harassment scenario:
(a) its clear association with an exceptional galaxy NGC~4254, located
1 Mpc from the center of Virgo and traveling at high speed;
(b) its overall length (242 kpc); (c) its modest gas mass ($6 \times 10^8$ \msun),
about 10\% of the total HI associated with NGC~4254; 
and (d) its roughly sinusoidal velocity field (right panel of Figure 1). 
Unlike the strongly deficient case of NGC~4388 (Oosterloo \& van Gorkom 2005), 
NGC~4254 is not HI poor, so that we can conclude it has not passed 
through the Virgo core in recent epochs. Its distance
of 1 Mpc from M87 is not coincidental: the galaxy density there is already
quite high, so that a close high speed encounter is not improbable.
A curious aspect of NGC~4254, cited in support of the
``dark galaxy'' hypothesis for Virgo~HI21 (Minchin \etal ~2005a), is its lack
of an obvious close companion which might be responsible for driving its
extraordinary spiral pattern. However, within the framework of the harassment
hypothesis, the  harassers need not be nearby: the collective cluster
potential, as we have seen, is centered 1 Mpc away and the galaxies with
which NGC 4254 may have had a close encounter would rapidly move far away
due to their high relative speed.
Moore \etal ~(1996a) further note that such harassment can result in 
disruption of the inner disk (Moore \etal ~1996a) as seen
in NGC~4254. 

Furthermore, the broad velocity width ($\sim 220$ \kms) of the section 
of the NGC~4254 HI tail identified as Virgo~HI21 by Minchin \etal ~(2005a,b) 
also led those authors to argue that the feature must be a massive rotating ``dark'' 
galaxy. As Bekki, Koribalski \& Kilborn (2005) have shown, gravitational
forces are more likely to produce broad velocity widths and large velocity
gradients than is ram pressure stripping. The full view of the tail provided
by ALFALFA shows the width of Virgo~HI21 to be not necessarily due to galaxy--scale
rotation, but rather a part of a larger scale velocity field.

In the harassment interpretation, the velocity field along the stream 
suggests a timing argument. We hypothesize that the roughly 
full cycle, sinusoidal velocity signature 
apparent in the right panel of Figure 1 reflects the coupling of 
the spin and tidal motions. Because gas is stripped more readily from that 
portion of the galaxy whose rotation aligns with the direction of the tidal
force, we assume that the HI present 
in the stream was removed from the edge of the outer disk of NGC~4254.
The HI diameter derived from the VLA maps is $\sim 7.7$\arcmin ~(Cayatte
\etal ~1994; Phookun, Vogel \& Mundy 1993), yielding an outer radius
of 18.5 kpc. Adopting the observed flat rotation curve with a
velocity in the outer portions of $\sim 150$ \kms ~(Guhathakurta
\etal ~1988; Phookun, Vogel \& Mundy 1993), the rotation period 
at a radius of 18.5 kpc corresponds to $\sim 7.5 \times 10^8$ years, which 
is comparable with the cluster crossing time. NGC~4254 is not a
particularly massive system: the mass enclosed within an 18.5 kpc radius 
is of order $10^{11}$ \msun; the Virgo cluster mass within a $6^\circ$ radius
is estimated to be $4.0\pm1.0\times 10^{14}$ \msun ~(Hoffman, Olson \& Salpeter
1980). Thus tidal forces due to high speed encounters and/or the collective 
cluster potential could easily strip the outer gas layers of NGC~4254 in a 
prograde orbit that passes within 1 Mpc of the cluster center.

Based on the discovery of relatively long stellar streams visible in very 
deep optical images of Virgo (Mihos \etal ~2005), Centaurus and Coma,
Calc\'aneo-Rold\'an \etal ~(2000) and Rudick, Mihos \& McBride (2006) 
have shown that such structures may arise when a relatively luminous spiral 
galaxy suffers tidal shocking on a prograde encounter deep within the 
cluster potential. The NGC~4254 event appears to be of a milder nature
than those illustrated in those works, presumably because NGC~4254 has
not plunged very deep into the cluster core. A stellar stream may be present
but would be harder to discern in this case, given the significantly
smaller optical light radius of the galaxy and the consequent rarefaction
of the stellar density outside 18.5 kpc.

We propose that the most likely interpretation of the galaxy's asymmetry,
its very extended HI tail and the origin of Virgo~HI21 is an on-going
process of galaxy harassment, resulting from the high speed gravitational
perturbations experienced by NGC~4254 as it enters the Virgo cluster. 

{\it Note added in proof:} We have recently learned that Bournaud \& Duc
(2007, in preparation) have modeled the formation of a tidal tail in
NGC~4254, with the characteristics described in this work,
resulting from a very high speed encounter with another cluster
galaxy now far removed from NGC~4254.

\vskip 0.3in

RG and MPH acknowledge the partial support of NAIC as Visiting Scientists during
the period of this work. This work has been supported by NSF grants AST--0307661,
AST--0435697, AST--0347929, AST--0407011, AST--0302049;
and by a Brinson Foundation grant.

\clearpage

%FIGURE 1
\begin{figure}[]
%\figurenum{1}
\epsscale{1.}
\plotone{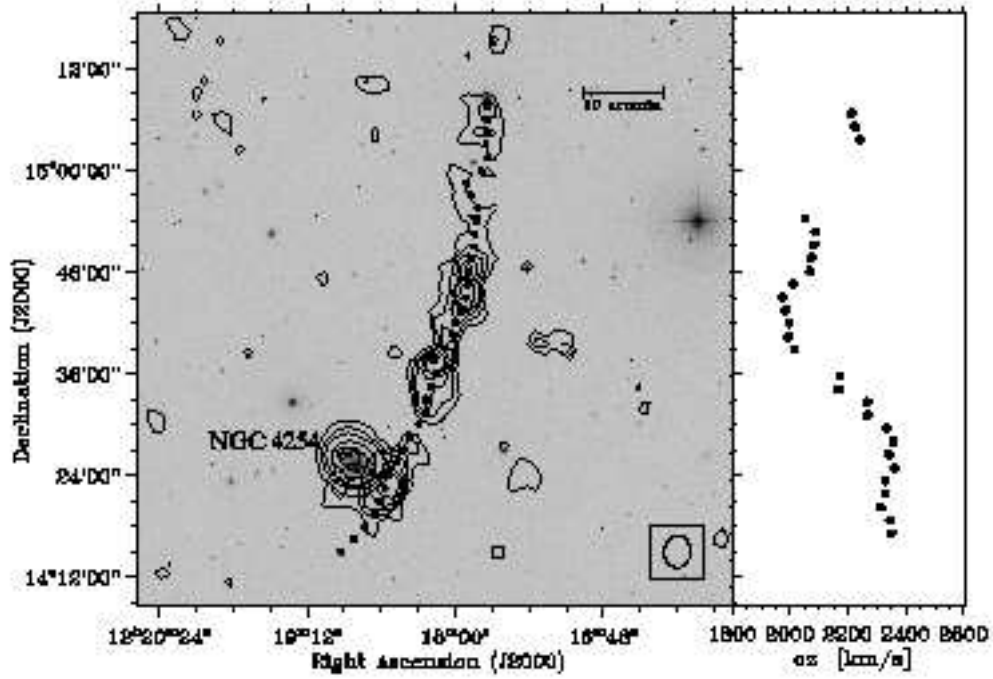}
\caption{Left: HI flux contours extracted from the ALFALFA
survey dataset, which mapped the full field represented
in the image, superposed on a DSS2 Blue image.  The 36 filled dots
indicate the locations of beam centers for the successive LBW 
observations.  The contours centered on NGC 4254 are at 10, 15, 20, 30, and 40
 Jy beam$^{-1}$ \kms, integrated from 2259 to 2621 \kms.
The contours for the HI stream are at 0.35, 0.52, 0.70, 0.87, and 1.0 Jy beam$^{-1}$ \kms, 
integrated from 1946 to 2259 \kms. Note the difference in
dynamic range of contours for the galaxy and the stream, selected for viewing ease.
The 3\arcmin ~circle in mid stream indicates the position of Virgo~HI21 reported by Minchin \etal (2005a).
The ellipse on the bottom right indicates the size of the Arecibo beam.
Right: The velocity
of the HI emission peak as seen in LBW pointings. Some of the LBW spectra yielded poor baselines
or poor peak definition.
}

\label{fig1}
\end{figure}

\end{document}